\documentclass[twoside]{mfa}

\usepackage{amssymb,amsfonts}
\renewcommand\volume{{\bf 8} (2019)}        

\usepackage[active]{srcltx}
\usepackage{graphicx}

\newcommand{\lteam}{\lambda}

\newcommand{\ppair}{\theta}

\newcommand{\wwin}{w}

\newcommand{\nnum}{n}

\newcommand{\num}{n}

\newcommand{\win}{v}

\newcommand{\nt}{t}
\newcommand{\ML}[1]{\widehat{#1}}
\newcommand{\MP}[1]{\widetilde{#1}}
\newcommand{\pdf}{f}
\newcommand{\pmf}{p}
\newcommand{\prob}{P}

\newcommand{\sref}[1]{Sec.~\ref{#1}}

\newcommand{\fref}[1]{Fig.~\ref{#1}}
\newcommand{\tref}[1]{Table~\ref{#1}}

\DeclareMathOperator{\logistic}{logistic}
\newcommand{\badgaussdist}{9.072}
 \newcommand{\elfCrKRACH}{415.3}
\newcommand{\elfQnKRACH}{93.30}
\newcommand{\elfCrQnKRACHpct}{81.7}
\newcommand{\elfCrQngausspct}{80.0}
\newcommand{\elfCrQnavgpcta}{80.0}
\newcommand{\elfCrQnavgpctb}{80.1}
\newcommand{\elfCrQnavgpctc}{80.0}
\newcommand{\elfCrQnavgpctd}{79.9}
\newcommand{\elfCrQnmcpcta}{80.2}
\newcommand{\elfCrQnmcpctb}{80.0}
\newcommand{\elfCrQnmcpctc}{79.7}
\newcommand{\elfCrQnmcpctd}{79.8}
\newcommand{\elfCrQnavgimppcta}{81.8}
\newcommand{\elfCrQnavgimppctb}{81.7}
\newcommand{\elfCrQnavgimppctc}{81.4}
\newcommand{\elfCrQnavgimppctd}{81.5}
\newcommand{\elfCrQnmcimppcta}{81.8}
\newcommand{\elfCrQnmcimppctb}{81.6}
\newcommand{\elfCrQnmcimppctc}{81.8}
\newcommand{\elfCrQnmcimppctd}{81.9}
\newcommand{\elfCrQnKRACHseriespct}{91.1}
\newcommand{\elfCrQngaussseriespct}{88.2}
\newcommand{\elfCrQnavgseriespcta}{88.2}
\newcommand{\elfCrQnavgseriespctb}{88.3}
\newcommand{\elfCrQnavgseriespctc}{88.1}
\newcommand{\elfCrQnavgseriespctd}{88.0}
\newcommand{\elfCrQnmcseriespcta}{88.3}
\newcommand{\elfCrQnmcseriespctb}{88.5}
\newcommand{\elfCrQnmcseriespctc}{87.7}
\newcommand{\elfCrQnmcseriespctd}{87.9}
\newcommand{\elfCrQnavgimpseriespcta}{89.8}
\newcommand{\elfCrQnavgimpseriespctb}{89.8}
\newcommand{\elfCrQnavgimpseriespctc}{89.5}
\newcommand{\elfCrQnavgimpseriespctd}{89.6}
\newcommand{\elfCrQnmcimpseriespcta}{89.8}
\newcommand{\elfCrQnmcimpseriespctb}{89.9}
\newcommand{\elfCrQnmcimpseriespctc}{89.6}
\newcommand{\elfCrQnmcimpseriespctd}{89.7}
\newcommand{\elfmaxweighta}{0.00399}
\newcommand{\elfmaxweightb}{0.00686}
\newcommand{\elfmaxweightc}{0.00488}
\newcommand{\elfmaxweightd}{0.00509}
 
\numberwithin{equation}{section}

\begin{document}

\title[Prediction in College Hockey using Bradley-Terry]
{Prediction and Evaluation in College Hockey
  using the Bradley-Terry-Zermelo Model}

\author[J.\ T.\ Whelan]{John T.\ Whelan}
\address{School of Mathematical Sciences and Center for Computational Relativity and Gravitation, Rochester Institute of Technology, 85 Lomb Memorial Drive, Rochester, New York 14623, USA \\ and Institute for Theoretical Physics, Goethe University Frankfurt, Max-von-Laue Str.\ 1, D-60438 Frankfurt am Main, Germany}
\email{jtwsma@rit.edu}

\author[A.\ Wodon]{Adam Wodon}
\address{College Hockey News}
\email{adamw@collegehockeynews.com}
\urladdress{https://www.collegehockeynews.com/}
 \date{\commitDATE}


{62F07}      

\begin{abstract}
  We describe the application of the Bradley-Terry model to NCAA Division I Men's Ice Hockey.  A Bayesian construction gives
  a joint posterior probability distribution for the log-strength
  parameters, given a set of game results and a choice of prior
  distribution.  For several suitable choices of prior, it is
  straightforward to find the maximum a posteriori point (MAP) and a
  Hessian matrix, allowing a Gaussian approximation to be constructed.
  Posterior predictive probabilities can be estimated by 1) setting
  the log-strengths to their MAP values, 2) using the Gaussian
  approximation for analytical or Monte Carlo integration, or 3)
  applying importance sampling to re-weight the results of a Monte
  Carlo simulation.  We define a method to evaluate
  any models which generate predicted
  probabilities for future outcomes, using the Bayes factor given the
  actual outcomes, and apply it to NCAA tournament results.  Finally,
  we describe an on-line tool which currently estimates probabilities
  of future results using MAP evaluation and describe how it can be
  refined using the Gaussian approximation or importance sampling.
\end{abstract}

\newtheorem{theorem}{Theorem}[section]
\newtheorem{corollary}[theorem]{Corollary}
\newtheorem{lemma}[theorem]{Lemma}
\newtheorem{proposition}[theorem]{Proposition}

\theoremstyle{definition}
\newtheorem{definition}[theorem]{Definition}
\newtheorem{problem}[theorem]{Problem}
\newtheorem{example}[theorem]{Example}
\newtheorem{remark}[theorem]{Remark}

\numberwithin{equation}{section}

\maketitle

\section{Introduction}

Sporting events, specifically games between pairs of teams, are a form
of paired comparison experiment, where one team (the winner) is chosen
over the other (the loser).  The Bradley-Terry model associates the
probability of the outcome of each paired comparison with the inherent
strengths of each team.  Estimates of these strengths can be used to
construct a rating system which allows the ranking of teams based on
game outcomes when imbalances in strength of schedule make simple
winning percentage (fraction of games won) an unfair basis for
ranking.  Because the strength parameters are also associated with the
probabilities of game results, they can be used to predict the outcome
of future games.  This paper considers the application of this
technique to NCAA Division I Men's Ice Hockey, the highest level of
college hockey competition in the United States.

The paper is organized as follows: In \sref{s:intro-bt} we define the
Bradley-Terry model and its use in constructing posterior estimates of
team strengths.  In \sref{s:intro-collegehockey} we describe the
particulars of the college hockey season and postseason.  In
\sref{s:probabilities} we describe several methods for estimating the
posterior predictive probabilities of the outcome of future games.  In
\sref{s:applications}, we describe two applications of these
techniques: the evaluation of the model constructed from the regular
season by use of a Bayes factor associated with NCAA tournament
results; and the \textit{Pairwise Probability Matrix}, used during the
season to construct predicted probabilities for a team's ranking
associated with the NCAA tournament selection criteria.

\subsection{The Bradley-Terry Model}
\label{s:intro-bt}

Given a set of $\nt$ teams, the Bradley-Terry-Zermelo model
\cite{BRADLEY01121952,Zermelo1929} associates with each team
$i=1,\ldots,\nt$ a log-strength parameter $\lteam_i$, and defines
the probability of team $i$ winning a given game with team $j$ such
that the odds ratio is the ratio of their strengths, i.e., the
probability is
\begin{equation}
  \ppair_{ij} = \frac{e^{\lteam_i}}{e^{\lteam_i}+e^{\lteam_j}}
  = \logistic(\lteam_i-\lteam_j)
\end{equation}
In this paper, we will work in terms of the log-strength
$\lteam_i\in(-\infty,\infty)$.

Given a series of games among the teams in which a pair of teams $i,j$
play $\nnum_{ij}=\nnum_{ji}$ times, the Bradley-Terry model defines a
probability for a set of outcomes $D$ which includes $\wwin_{ij}$ wins
(and $\nnum_{ij}-\wwin_{ij}$ losses) for team $i$ against team $j$:
\begin{equation}
  \prob(D|\{\lteam_i\}) = \prod_{i=1}^{\nt}\prod_{j=1}^{\nt} \ppair_{ij}^{\wwin_{ij}}
  = \prod_{i=1}^{\nt}\prod_{j=i+1}^{\nt}
  \ppair_{ij}^{\wwin_{ij}}(1-\ppair_{ij})^{\nnum_{ij}-\wwin_{ij}}
\end{equation}
Note that if the order of the outcomes of games between pairs of teams
is ignored, the sampling distribution for $\wwin_{ij}$ is
\begin{equation}
  \pmf(\{\wwin_{ij}\}|\{\lteam_i\})
  = \prod_{i=1}^{\nt}\prod_{j=i+1}^{\nt}\binom{\nnum_{ij}}{\wwin_{ij}}
  \ppair_{ij}^{\wwin_{ij}}(1-\ppair_{ij})^{\nnum_{ij}-\wwin_{ij}}
\end{equation}
but the inferences about parameters $\{\lteam_i\}$ are unchanged.

The log-likelihood can be written in terms of the total number of wins
$\win_i=\sum_{j=1}^{\nt} \wwin_{ij}$ as
\begin{equation}
  \ln \prob(D|\{\lteam_i\}) = \sum_{i=1}^{\nt} \win_i\lteam_i
  -\frac{1}{2}\sum_{i=1}^{\nt}\sum_{j=1}^{\nt} \nnum_{ij}
  \ln\left(e^{\lteam_i}+e^{\lteam_j}\right)
\end{equation}
so that the maximum likelihood equations are
\begin{equation}
  \win_i
  = \sum_{j=1}^{\nt} \nnum_{ij}
  \frac{e^{\ML{\lteam}_i}}{e^{\ML{\lteam}_i}+e^{\ML{\lteam}_j}}
  = \sum_{j=1}^{\nt} \nnum_{ij}\ML{\ppair}_{ij}
\end{equation}
The maximum-likelihood log-strengths $\{\ML{\lteam}_i\}$ are those for
which the predicted number of wins
$\sum_{j=1}^{\nt} \nnum_{ij}\ML{\ppair}_{ij}$ for each team equals the
actual number $v_i$.  They can be found, e.g., by Ford's method, in
which one iterates the equation\cite{Ford:1957}
\begin{equation}
  \ML{\lteam}_i = \ln
  \left(
    \win_i\left/
      \sum_{j=1}^{\nt}
      \frac{\nnum_{ij}}{e^{\ML{\lteam}_i}+e^{\ML{\lteam}_j}}
    \right.
  \right)
\end{equation}
Because the maximum-likelihood equations depend only on the
differences $\ML{\lteam}_i-\ML{\lteam}_j$, the estimates
$\{\ML{\lteam}_i\}$ are defined only up to an overall additive
constant.

Given a prior distribution $\pdf(\{\lteam_i\}|I)$ for the
log-strengths, the posterior distribution given the game results $D$
will be
\begin{equation}
  \pdf(\{\lteam_i\}|D,I) \propto \prob(D|\{\lteam_i\})\,\pdf(\{\lteam_i\}|I)
\end{equation}
The maximum a posteriori (MAP) estimates $\{\MP{\lteam_i}\}$ of the
log-strengths will be the solution to
\begin{equation}
  \win_i
 + \left.
    \frac{\partial}{\partial\lteam_i} \ln\pdf(\{\lteam_j\}|I)
  \right\rvert_{\{\lteam_j=\MP{\lteam}_j\}}
  = \sum_{j=1}^{\nt} \nnum_{ij}\MP{\ppair}_{ij}
\end{equation}
For the sake of mathematical simplicity, we will often use the Haldane
prior\footnote{So named because the marginal prior distribution
  for any $\ppair_{ij}$ will follow the Haldane prior
  \cite{Haldane1932,Jeffreys1939}, which is the limit of a
  $\text{Beta}(\alpha,\beta)$ distribution as
  $\alpha,\beta\rightarrow 0$.}
\begin{equation}
  \pdf(\{\lteam_i\}|I_0) = \text{constant}
\end{equation}
This is an improper prior
but can be formally understood as the limiting form of a family of
normalized priors.

Other convenient families of priors\cite{Whelan2017} are the
generalized logistic prior
\begin{equation}
  \pdf(\{\lteam_i\}|I_{\eta})
  = \prod_{i=1}^{\nt} \frac{\Gamma(2\eta)}{[\Gamma(\eta)]^2}
\frac{1}{(1+e^{\lteam_i})^\eta(1+e^{-\lteam_i})^\eta}
\end{equation}
and the Gaussian prior
\begin{equation}
  \pdf(\{\lteam_i\}|I_{\sigma})
  = \prod_{i=1}^{\nt} \frac{1}{\sigma\sqrt{2\pi}}
  \,\exp\left(
    -\frac{(\lteam_i)^2}{2\sigma^2}
  \right)
\end{equation}
The Haldane prior is the limit of the generalized logistic prior as
$\eta\rightarrow 0$ and the Gaussian prior as
$\sigma\rightarrow\infty$.

Since the generalized logistic prior has
\begin{equation}
\ln\pdf(\{\lteam_j\}|I_{\eta})
=  \sum_{j=1}^{\nt}\left[-\eta\ln(1+e^{-\lteam_j}) -\eta\ln(1+e^{\lteam_j})\right]
+ \text{constant}
\end{equation}
and
\begin{equation}
\frac{\partial}{\partial\lteam_i} \ln\pdf(\{\lteam_j\}|I_{\eta})
= \frac{\eta\,e^{-\lteam_i}}{1+e^{-\lteam_i}}
-\frac{\eta\,e^{\lteam_i}}{1+e^{\lteam_i}}
= \eta(1-2\ppair_{i0})
\end{equation}
where $\ppair_{i0}=\logistic(\lteam_i)$ is the probability that team
$i$ would win a game against a team with log-strength zero.  This
means that the MAP equations with the generalized logistic prior are
\begin{equation}
  \eta + \win_i
  = 2\eta\MP{\ppair}_{i0}
  + \sum_{j=1}^{\nt} \nnum_{ij}\MP{\ppair}_{ij}
\end{equation}
This is just the same as we'd obtain from the maximum-likelihood
equations after the addition of $2\eta$ ``fictitious games'' against a
team with log-strength zero, half wins and half losses, for each team.
As such, the MAP equations can be solved by a straightforward
extension of Ford's method.

With the Gaussian prior, the MAP equations become
\begin{equation}
  \win_i
  = \frac{\MP{\lteam}_i}{\sigma^2}
  + \sum_{j=1}^{\nt} \nnum_{ij}\MP{\ppair}_{ij}
\end{equation}
Note that these cannot in general be solved by iterating
\begin{equation}
  \MP{\lteam}_i
  =
  \ln
  \left(
  \frac{
    \win_i - \frac{\MP{\lteam}_i}{2\sigma^2}
  }{
    \sum_{j=1}^{\nt}
    \frac{\nnum_{ij}}{e^{\MP{\lteam}_i}+e^{\MP{\lteam}_j}}
  }
  \right)
\end{equation}
as suggested in \cite{Phelan2017} because, for small values of
$\sigma$, the argument of the logarithm may become negative.

\subsection{College Hockey}
\label{s:intro-collegehockey}

The NCAA (National Collegiate Athletic Association) Men's Division I
Ice Hockey competition consists, at present, of 60 teams which play
approximately 30 to 40 games each during the season.  At the end of the
season, 16 teams are selected (the champions of six conference
tournaments, plus an additional 10 teams chosen according to a set of
selection criteria related to the outcomes of their games) to play in
a single-elimination tournament to determine the national champion.
As the games during the season are played within six conferences (with
one team currently competing as an independent), with additional
non-conference games, in-season tournaments, and conference playoff
tournaments, teams will typically face schedules of differing
strengths.  Rating systems have thus been devised to evaluate their
game results more fairly than would be possible by simply comparing
winning percentages (fraction of games won).  The selection criteria
for the NCAA tournament are of this sort, notably the ratings
percentage index (RPI) which combines a team's winning percentage with
average winning percentages of its opponents and opponents' opponents.
The maximum-likelihood Bradley-Terry strengths are also used, under
the name ``Ken's Ratings for American College Hockey''
(KRACH)\cite{KRACH1993}.

In addition to a win or a loss, some college hockey games can end in a
tie.  In computing NCAA selection criteria, a team which wins (whether
in regulation play or overtime) is awarded 2 points, a team which
loses receives 0 points, and if a game ends in a tie (after overtime),
each team receives 1 point.  (Penalty shootouts, which may occur after
a tie in some competitions, are not considered for NCAA selection
purposes.)  In principle, one could use an extension of the
Bradley-Terry model with an additional parameter or parameters
accounting for the probability of ties.
\cite{Rao1967,Davidson1970,Joe1990} However, this is complicated by
the fact that some college hockey games can end in ties, while others
(mostly conference playoff and NCAA tournament games, but also some
games in in-season invitational tournaments) continue in overtime
until a winner is decided.  Rather than keep track of the two sorts of
games, in this work we perform all computations with ties contributing
$0.5$ to the win total and $0.5$ to the loss total for each team.
While this introduces a conceptual inconsistency (since the formulas
were derived without consideration for the possibility of ties), it
poses no impediment to the calculations, and ties are rare enough that
no pathological conclusions have yet been encountered.

\section{Posterior Predictive Probabilities}
\label{s:probabilities}

In this paper, we are interested in the calculation of posterior
predictive probabilities.  I.e., given a set of results $D$ and a
choice of prior $I$, so that the Bayesian Bradley-Terry model produces
a posterior pdf $\pdf(\{\lteam_i\}|D,I)$, what is the probability that
some future games will have outcome $O$?  This outcome may be a
specified set of wins and losses, but it may also be a more
coarse-grained alternative, such as that a particular team is chosen
for the tournament field by winning its conference championship game
or finishing the season well enough according to the selection
criteria.  If $\prob(O|\{\lteam_i\})$ is the probability of that outcome
using the Bradley-Terry model with log-strengths $\{\lteam_i\}$ (which
might itself be calculated according to some non-trivial summation
technique) then the posterior predictive probability for $O$ given the
previous results $D$ and prior information $I$ is obtained by
marginalizing over the log-strengths:
\begin{equation}
  \label{e:postpred}
  \prob(O|D,I) = \int_{-\infty}^{\infty}\cdots \int_{-\infty}^{\infty}
  \prob(O|\{\lteam_i\})\,\pdf(\{\lteam_i\}|D,I)\, d^{\nt}\!\lteam
\end{equation}
While the integrand of \eqref{e:postpred} may be straightforward to
construct, the $\nt$-dimensional integral is in general impossible to
evaluate analytically and impractical to compute via direct numerical
integration.  One approach is to draw samples from the posterior
$\pdf(\{\lteam_i\}|D,I)$ via Markov Chain Monte Carlo methods, such as
the Hamiltonian Monte Carlo used in \cite{Phelan2017}.  In this paper,
we consider simpler techniques which apply approximation methods.

\subsection{MAP Evaluation}
\label{s:probabilities-MAP}

The simplest method is to use the Bradley-Terry model with the team
strength parameters set to their maximum a posteriori values
$\{\MP{\lteam}_i\}$, and evaluate $\prob(O|\{\MP{\lteam}_i\})$.  This is
equivalent to replacing the full posterior $\pdf(\{\lteam_i\}|D,I)$
with a $\nt$-dimensional delta function at the MAP point.  While this
is convenient, it is clearly an oversimplification, since it fails to
account for the posterior uncertainty in the team strengths.

\subsection{Gaussian Approximation}
\label{s:probabilities-gauss}

One way to quantify the uncertainty in the posterior, and obtain a
better approximation, is to Taylor expand the log-posterior
$\ln\pdf(\{\lteam_i\}|D,I)$ about the MAP point and obtain a Gaussian
approximation
\begin{equation}
  g(\{\lteam_i\}|D,I)
  =\text{const}\times \exp\left(
    -\frac{1}{2}\sum_{i=1}^{\nt}\sum_{j=1}^{\nt}
    \left[\lteam_i-\MP{\lteam}_i\right]
    H_{ij}
    \left[\lteam_j-\MP{\lteam}_j\right]
  \right)
\end{equation}
where the Hessian matrix $\{H_{ij}\}$ has elements
\begin{equation}
  \begin{split}
    H_{ij} &= -\left.
      \frac{\partial^2}{\partial\lteam_i\,\partial\lteam_j}
      \ln\pdf(\{\lteam_i\}|D,I)
    \right\rvert_{\{\lteam_k=\MP{\lteam}_k\}}
    \\
    &= -\nnum_{ij}\MP{\ppair}_{ij}\MP{\ppair}_{ji}
    + \delta_{ij}\sum_{k=1}^{\nt}\MP{\ppair}_{ik}\MP{\ppair}_{ki}
    - \left.
      \frac{\partial^2}{\partial\lteam_i\,\partial\lteam_j}
      \ln\pdf(\{\lteam_i\}|I)
    \right\rvert_{\{\lteam_k=\MP{\lteam}_k\}}
  \end{split}
\end{equation}
For the Haldane prior, the last term vanishes; for the generalized
logistic prior it is
\begin{equation}
  -\left.
    \frac{\partial^2}{\partial\lteam_i\,\partial\lteam_j}
    \ln\pdf(\{\lteam_i\}|I_{\eta})
  \right\rvert_{\{\lteam_k=\MP{\lteam}_k\}}
  = \delta_{ij}\, \MP{\theta}_{i0}(1-\MP{\theta}_{i0})
\end{equation}
and for the Gaussian prior, it is
\begin{equation}
  -\left.
    \frac{\partial^2}{\partial\lteam_i\,\partial\lteam_j}
    \ln\pdf(\{\lteam_i\}|I_{\sigma})
  \right\rvert_{\{\lteam_k=\MP{\lteam}_k\}}
  = \frac{\delta_{ij}}{\sigma^2}
\end{equation}

The Gaussian approximation is a multivariate Gaussian distribution
with mean $\{\MP{\lteam}_i\}$ and a variance-covariance matrix
$\{\Sigma_{ij}\}$ which is the matrix inverse of the Hessian
$\{H_{ij}\}$.  Given a suitable prior distribution, $\{H_{ij}\}$ is
invertible.  Using the Haldane prior produces a Hessian matrix
\begin{equation}
H_{ij}
= -\nnum_{ij}\ML{\ppair}_{ij}\ML{\ppair}_{ji}
+ \delta_{ij}\sum_{k=1}^{\nt}\ML{\ppair}_{ik}\ML{\ppair}_{ki}
= \sum_{k=1}^{\nt} h_k \ell^{(k)}_i \ell^{(k)}_j
\end{equation}
Where we have decomposed the Hessian matrix using its orthonormal
eigenvectors
$\sum_{i=1}^{\nt}\ell^{(k)}_i \ell^{(\ell)}_i=\delta_{k\ell}$,
$\sum_{j=1}^{\nt}H_{ij}\ell^{(k)}_j=h_k \ell^{(k)}_i$,
$h_i\le h_{i+1}$.  There will be at least one zero eigenvalue $h_1=0$,
corresponding to the eigenvector
$\{\ell^{(1)}_i\}=\{\frac{1}{\sqrt{\nt}},\frac{1}{\sqrt{\nt}},\ldots,\frac{1}{\sqrt{\nt}}\}$.
If all of the maximum-likelihood estimates
$\{\ML{\lteam}_i-\ML{\lteam}_j\}$ are finite and
well-determined\cite{Albert1984,Santner1986,ButlerWhelan},
which will nearly always be the case late in a season with as many
games as college hockey, the other eigenvalues
$\{h_i|i=2,\ldots,\nt\}$ will all be positive.
Since
the transformation $\lteam_i\rightarrow\lteam_i + a\ell^{(1)}_i$, for
any $a\in\mathbb{R}$,
doesn't change the probabilities $\prob(O|\{\lteam_i\})$, we can
replace the Gaussian approximate distribution, which leaves
$\sum_{i=1}^{\nt}\ell^{(1)}_i\lteam_i$ unspecified, and is therefore
unnormalizable, with one which fixes
$\sum_{i=1}^{\nt}\ell^{(1)}_i\lteam_i=0$.  This is a multivariate
Gaussian distribution whose mean is $\{\ML{\lteam}_i\}$ and whose
variance-covariance matrix is the Moore-Penrose
pseudo-inverse\cite{penrose_1955} of
$H_{ij}$, i.e.,
\begin{equation}
  \Sigma_{ij} = \sum_{k=2}^{\nt} \frac{\ell^{(k)}_i \ell^{(k)}_j}{h_k}
  \ ,
\end{equation}
defined such that
\begin{equation}
  \sum_{k=1}^{\nt} \Sigma_{ik}H_{kj} = \sum_{k=1}^{\nt} H_{ik}\Sigma_{kj}
  = \delta_{ij} - \ell^{(1)}_i \ell^{(1)}_j
  \ .
\end{equation}

Depending on the specifics of the outcome $O$, it may be possible to
evaluate the approximate integral
\begin{equation}
  \prob(O|D,I) \approx \int_{-\infty}^{\infty}\cdots \int_{-\infty}^{\infty}
  \prob(O|\{\lteam_i\})\,g(\{\lteam_i\}|D,I)\, d^{\nt}\!\lteam
\end{equation}
analytically using the Gaussian approximation.  More likely it will be
necessary to use a Monte Carlo technique, drawing $N$ samples
$\{\lteam_i^{(s)}\}$ from the multivariate Gaussian distribution
$N_{\nt}(\{\MP{\lteam}_i\},\{\Sigma_{ij}\})$ and estimating
\begin{equation}
  \label{e:gausssample}
  \prob(O|D,I) \approx \frac{1}{N} \sum_{s=1}^{N} \prob(O|\{\lteam_i^{(s)}\})
\end{equation}

\subsection{Importance Sampling}
\label{s:probabilities-importance}

\begin{figure}[t!]
  \centering
  \includegraphics[width=\textwidth]{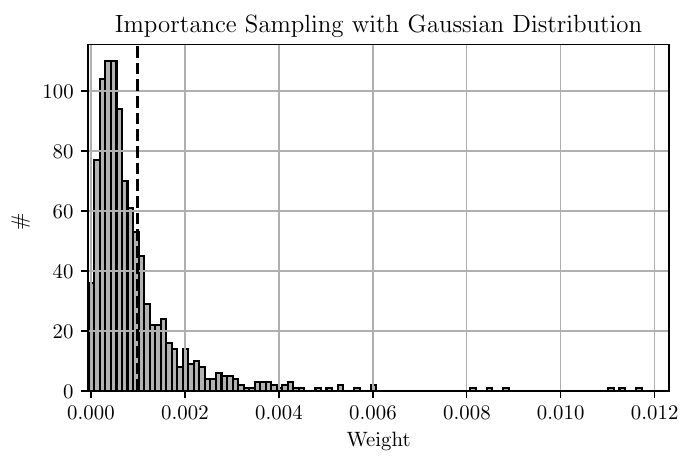}
  \caption{Histogram of weights in importance sampling using $1000$
    draws from a multivariate ($\nt=60$, one degenerate degree of
    freedom) Gaussian sampling distribution with to approximate the
    Bradley-Terry posterior, starting with the Haldane prior and the
    results of the 2018-2019 NCAA Division I Men's Ice Hockey season
    prior to NCAA tournament selection.  The dashed line indicates
    average weight of $0.001$.}
  \label{f:gaussimpratiohist}
\end{figure}
\begin{figure}[t!]
  \centering
  \includegraphics[width=\textwidth]{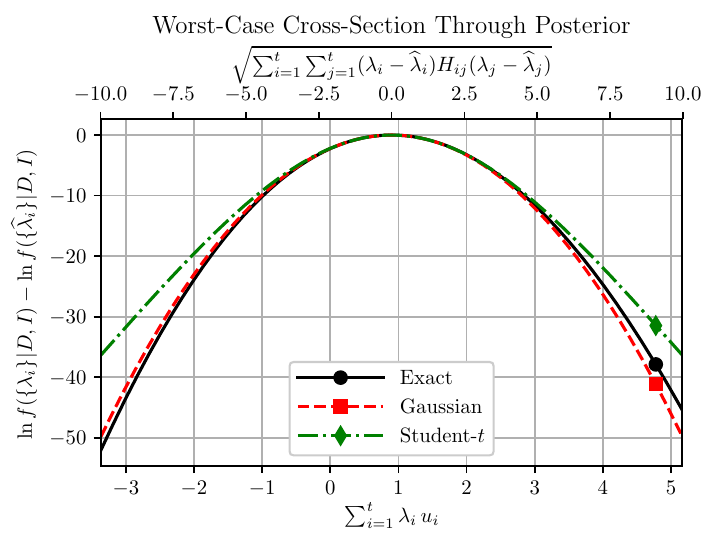}
  \caption{Cross-section (conditional distribution) of the log-posterior
    $\ln\pdf(\{\lteam_i^{(s)}\}|D,I)$ through the point with the highest
    weight in the Gaussian importance sampling.  While the
    multivariate Gaussian is a good approximation for some distance
    from the MAP point, the presence of $\nt-1=59$ meaningful
    parameters means that even seemingly large outliers can occur in
    the Gaussian sample}
  \label{f:crosssection}
\end{figure}
Since $g(\{\lteam_i\}|D,I)$ is only an approximation to
$\pdf(\{\lteam_i\}|D,I)$, a natural correction to
\eqref{e:gausssample} is to use importance sampling, weighting the
probability $\prob(O|\{\lteam_i^{(s)}\})$ coming from each Monte Carlo
draw by a factor
\begin{equation}
  w_s \propto
  \frac{\pdf(\{\lteam_i^{(s)}\}|D,I)}{g(\{\lteam_i^{(s)}\}|D,I)}
\end{equation}
so that
\begin{equation}
  \prob(O|D,I) \approx
  \frac{\sum_{s=1}^{N} w_s \prob(O|\{\lteam_i^{(s)}\})}
  {\sum_{s=1}^{N} w_s}
  \ .
\end{equation}
If there are values of $\{\lteam_i\}$ for which
\begin{equation}
  \frac{g(\{\lteam_i^{(s)}\}|D,I)}{\pdf(\{\lteam_i^{(s)}\}|D,I)}
  \gg \frac{1}{N}\sum_{s'=1}^N
  \frac{g(\{\lteam_i^{(s')}\}|D,I)}{\pdf(\{\lteam_i^{(s')}\}|D,I)}
  \ ,
\end{equation}
the importance sampling procedure may produce erratic results.  We
see this effect when we use the Gaussian approximation for importance
sampling.  We see in \fref{f:gaussimpratiohist} that a few outliers
produce large weight factors to try to adjust for the fact that the
tail of the posterior is longer than that of the approximate Gaussian
distribution.

This is illustrated in \fref{f:crosssection}, which shows a slice
through the log-posterior including the maximum-likelihood point
$\{\ML{\lteam}_i\}$ and the point $\{\lteam_i^{(\hat{s})}\}$ with the
largest importance sampling weight.  The lower $x$-axis shows the
projection of the vector $\{\lteam_i\}$ onto the unit vector $\{u_i\}$
defined by
\begin{equation}
  u_i = \frac{\lteam_i^{(\hat{s})}-\ML{\lteam}_i}
  {\sqrt{\sum_{j=1}^\nt (\lteam_j^{(\hat{s})}-\ML{\lteam}_j)^2}}
\end{equation}
The upper $x$-axis shows the normalized distance
\begin{equation}
  \sqrt{
    \sum_{i=1}^{\nt}\sum_{j=1}^{\nt}
    \left[\lteam_i-\ML{\lteam}_i\right]
    H_{ij}
    \left[\lteam_j-\ML{\lteam}_j\right]
  }
\end{equation}
\begin{figure}[t!]
  \centering
  \includegraphics[width=\textwidth]{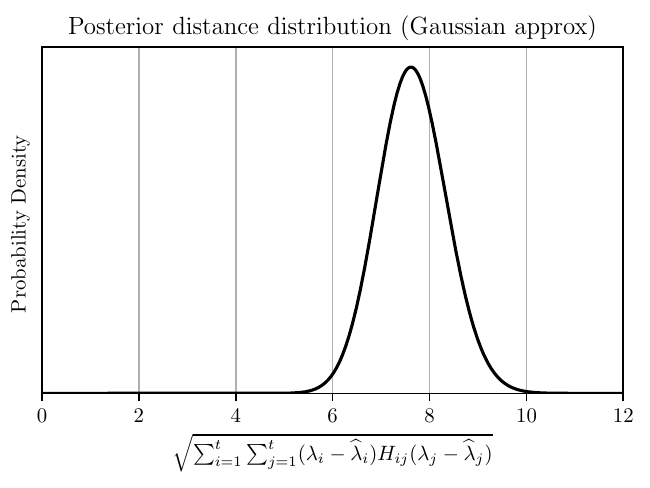}
  \caption{Marginal probability density function for the normalized
    distance from the maximum-likelihood point under multivariate
    Gaussian random sampling, which is a $\chi(59)$ random variable.
    Essentially all the probability weight is between 5 and 10 sigma
    from the ML point.}
  \label{f:chi59dens}
\end{figure}
from the maximum-likelihood point.  The largest importance sampling
rate occurs at a normalized distance of ${\badgaussdist}$-sigma from
the ML point.  This is not as large an outlier as it might seem.  With
$\nt-1=59$ meaningful parameters, the normalized distance in a
Gaussian Monte Carlo will be a chi-distributed random variable with
$\nt-1=59$ degrees of freedom, whose probability density function is
shown in \fref{f:chi59dens}.  It shows that the samples are
overwhelmingly likely to be found between 5 and 10 sigma from the ML
point.  This means that using a ``heavy-tailed'' distribution such as
the multivariate Student-$t$ distribution will not improve the
situation.  While the Student-$t$ distribution has more support at
large distances relative to the maximum-likelihood point, essentially
none of the random samples will be near the ML point.  Instead, the
differently-shaped tails of the $t$-distribution would cause points
less far from the ML point to be undersampled relative to points
farther out, producing larger outliers in the normalized weights.
This is illustrated in \fref{f:crosssection} for a multivariate
Student-$t$ distribution with $\nu=59$ degrees of freedom.  The
covariance matrix has been scaled up so that a one-dimensional
``slice'' through the maximum, i.e., the conditional distribution
\cite{Ding2016} is a Student-$t$ distribution with $\nu+(\nt-1)-1$
degrees of freedom and (pseudo-)inverse scale matrix $\{H_{ij}\}$.

Given that the departure of the posterior from normality (at least in
this example) is a matter of slight skewness than heavy tails, an
avenue for future exploration is importance sampling with a skew
distribution, as proposed in \cite{Swartz2005,WangSwartz}.

\section{Applications}
\label{s:applications}

\subsection{Evaluation via Bayes Factor}

\label{s:applications-bayes}

In \sref{s:probabilities}, we described methods to calculate or
approximate the probability of future outcomes $O$ using the
Bradley-Terry model.  We now describe a simple method for evaluating
any set of predictions.  Suppose $\prob(O|M_1,D,I)$ and $\prob(O|M_2,D,I)$ are
the probabilities assigned to a future outcome $O$ by two different
methods $M_1$ and $M_2$, given past results $D$ and any additional
information $I$.  (These should be defined so that
$\sum_O \prob(O|M,D,I)=1$ for any exhaustive set of mutually exclusive
outcomes $\{O\}$.)  A general method for comparing $M_1$ and $M_2$ is
the Bayes factor
\begin{equation}
  B_{12} = \frac{\prob(O|M_1,D,I)}{\prob(O|M_2,D,I)}
\end{equation}
which is the factor by which the posterior odds ratio for $M_1$ over
$M_2$ increases relative to the prior odds ratio:
\begin{equation}
  \frac{\prob(M_1|O,D,I)}{\prob(M_2|O,D,I)}
  = \frac{\prob(O|M_1,D,I)}{\prob(O|M_2,D,I)}
  \frac{\prob(M_1|I)}{\prob(M_2|I)}
  = B_{12}
  \frac{\prob(M_1|I)}{\prob(M_2|I)}
\end{equation}
We can apply this technique to any method of generating probabilities
for future outcomes of hockey games (not just Bradley-Terry).  We can
think of the results $D$ as ``training data'' and the outcome $O$ as
describing the ``evaluation data'' of the rest of the games.  We
consider a straightforward example, where the training data are the
games of each season prior to tournament selection and the evaluation
data are the NCAA tournament games, with $O$ being the actual sequence
of results which occurred.  Note that for this evaluation calculation,
we don't actually need to know $\prob(O|M,D,I)$ for each possible outcome,
only for the exact sequence of results which occurred.  For
convenience, we compare each model to a ``tossup model'' $M_0$ in
which each team is assigned a 50\% chance to win each game, for which
$\prob(O|M_0,D,I)=2^{-n_O}$ where $n_O$ is the number of games in the
evaluation data set.  Evidently $B_{12}=B_{10}/B_{20}$.

\begin{figure}[t!]
  \label{f:KRACH2019BF}
  \centering
  \includegraphics[width=\textwidth]{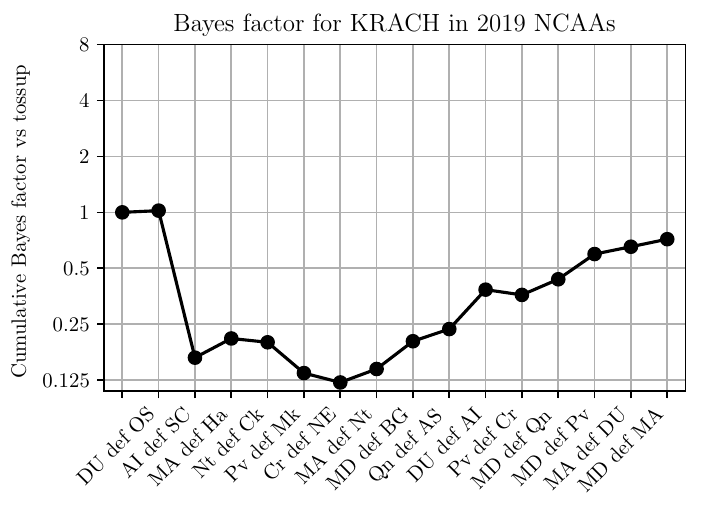}
  \caption{Evolution of the Bayes factor for the predictions of the
    maximum likelihood Bradley Terry model (KRACH) over the 2019 NCAA
    tournament.
}
\end{figure}

If $M_{\text{mle}}$ is the MAP evaluation method of
\sref{s:probabilities-MAP}, in which all probabilities are
independently assigned using the maximum-likelihood Bradley-Terry
estimates (the KRACH ratings),
\begin{equation}
  B_{\text{mle}0} = \prod_{g=1}^{n_O} 2\ML{\ppair}_{w_gl_g}
\end{equation}
where $w_g$ is the winner and $l_g$ the loser of game $g$.  So we see
that for each game predicted ``correctly'' (winner assigned a greater
than 50\% probability), the Bayes factor increases by a factor of up
to $2$.  However, for each game predicted ``incorrectly'' (winner
assigned a less than 50\% probability), the Bayes factor decreases.
If a result occurs which the model considered impossible, the Bayes
factor is zero.  We can illustrate this with the results of the 2019
NCAA tournament, in \fref{f:KRACH2019BF} We see that the Bayes factor
using all the results of the tournament is actually slightly lower
than 1.  This is because the upset of American International College
defeating St.~Cloud State was such a surprise according to the model.

\begin{figure}[t!]
  \centering
  \includegraphics[width=\textwidth]{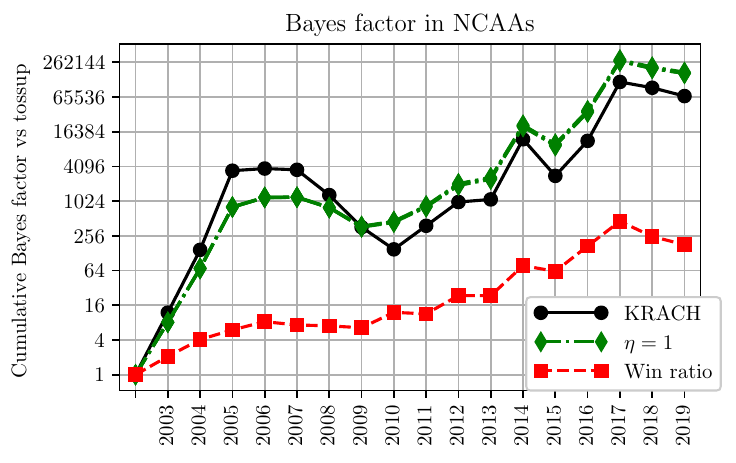}
  \caption{Evolution of the Bayes factor for the predictions of the
    three models over the NCAA tournament since the 16-team format was
    introduced in 2003: the maximum likelihood Bradley Terry model
    (KRACH), a model with a generalized logistic prior with $\eta=1$
    and a na\"{\i}ve model where the probability for a team to win a
    game is proportional to the square root of its win ratio (wins
    divided by losses) without regard to strength of schedule.}
  \label{f:NCAABF}
\end{figure}

If we compute the Bayes factor using the predictions and outcomes of
multiple NCAA tournaments (using the game results from each season to
produce probabilities for that season's tournament), we begin to see
distinctions between models.  In \fref{f:NCAABF} we plot the evolution
of this cumulative Bayes factor over the NCAA tournaments from 2003
(the first year of the current 16-team format) to 2019.  In addition
to the maximum likelihood/KRACH model, we plot the Bayes factor for a
model with a generalized logistic prior with $\eta=1$ (estimated using
the Gaussian approximation and 20,000 Monte Carlo draws)\footnote{Note
  that \fref{f:NCAABF} contains the results of four different Monte
  Carlo simulations (each with 20,000 draws for each season) plotted
  on top of one another, to illustrate that the integrals of the
  Gaussian-approximated posterior have been estimated accurately.  If
  a similar exercise is performed with Gaussian importance sampling,
  the four simulations give vastly different posterior predictive
  probabilities, indicating the algorithm is not stable enough to
  estimate the small probability associated with one particular
  sequence of results.}, along with a
simple model based on the win ratios $\win_i/(\num_i-\win_i)$ for each
team, where the probability that team $i$ will beat team $j$ is
assumed to be $\ppair^{\text{wr}}_{ij}$, where
\begin{equation}
  \frac{\ppair^{\text{wr}}_{ij}}{\ppair^{\text{wr}}_{ji}}
  = \frac{\ppair^{\text{wr}}_{ij}}{1-\ppair^{\text{wr}}_{ij}}
  = \sqrt{\frac{\win_i}{\num_i-\win_i}\frac{\num_j-\win_j}{\win_j}}
\end{equation}

We can see that 17 tournaments of 15 games each are enough to show
that the Bradley-Terry model is clearly preferred to the model using
win ratios without including strength of schedule, which is in turn
better than declaring each game a tossup.  It is not enough, however,
to establish a preference between the Haldane and generalized logistic
priors, although their predictions have not always been identical.

\subsection{The Pairwise Probability Matrix}

\label{s:applications-matrix}

The Pairwise Probability Matrix \cite{Wodon2017} is a tool to predict
the probability that each team will make the NCAA tournament.  It
typically runs with a few weeks remaining before the end of the
conference tournaments and the selection of the tournament field.  In
its current configuration (2018-2019 season), it takes a set of
Bradley-Terry log-strengths $\{\lteam_i\}$ and estimates the
probability $\prob(O|\{\lteam_i\})$ for an outcome $O$ (typically a team
being selected for the NCAA tournament) as follows:
\begin{figure}[t!]
  \centering
  \includegraphics[width=\textwidth]{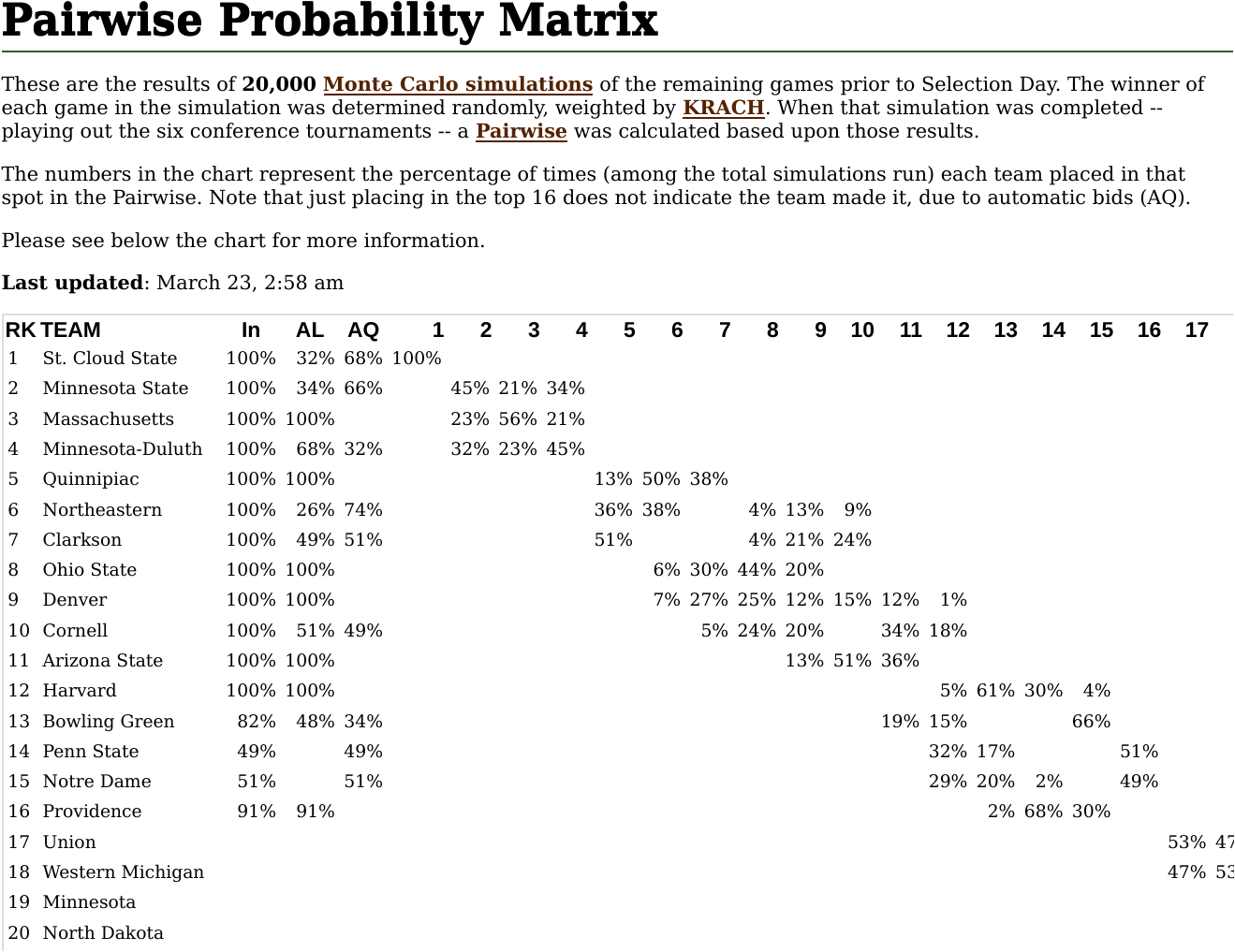}
  \caption{An excerpt of the Pairwise Probability Matrix display,
    displaying estimated probabilities entering the final day of games
    (2019 March 23) before NCAA tournament selection.  Excerpt from
    \texttt{https://www.collegehockeynews.com/ratings/probabilityMatrix.php}}
  \label{f:probabilitymatrix}
\end{figure}
\begin{enumerate}
\item A set of $N=20,\!000$ Monte Carlo trials are run.  In each trial:
  \begin{enumerate}
  \item The remaining games of the season are simulated; in each game,
    a winner is randomly chosen according to the probability predicted
    by the Bradley-Terry model.  For instance, if team $i$ plays team
    $j$, the probability that $i$ will win is modelled as
    $\theta_{ij}=\logistic(\lteam_i-\lteam_j)$, and team $i$ is
    assigned as the winner if a $\text{Uniform}(0,1)$ random draw is
    less than $\theta_{ij}$.
  \item The games to be played are not pre-determined, but may depend
    on the results of other games earlier in the simulation (e.g., the
    loser of a game may be eliminated from a conference tournament).
  \item When all the games have been simulated, teams are evaluated
    according to the NCAA selection criteria, including an ordering
    based on pairwise comparisons, and automatic qualification for the
    winners of conference tournaments.
  \end{enumerate}
\item The probability of an outcome $O$ is approximated as the
  fraction of Monte Carlo simulations in which it occurs.
\end{enumerate}
At present, the ratings used are the maximum likelihood estimates
$\{\ML{\lteam_i}\}$, expressed as KRACH ratings
$\{100\,e^{\ML{\lteam_i}}\}$, so that the probability of a future
outcome given past game results $D$ is approximated as in
\sref{s:probabilities-MAP}:
\begin{equation}
  \prob(O|D,I) \approx \prob(O|\{\ML{\lteam}_i\})
  \approx \frac{1}{N}\sum_{s=1}^N I^{(s)}(O)
  \ ,
\end{equation}
where $I^{(s)}(O)=1$ if $O$ occurs in Monte Carlo trial $s$ and $0$ if
not.  An excerpt of a typical display, taken before the final day of
games of the 2018-2019 season, is shown in \fref{f:probabilitymatrix}.

\subsubsection{Shortcomings of the MLE Probabilities}

\label{s:applications-matrix-CrQn}

\begin{figure}[t!]
  \centering
  \includegraphics[width=\textwidth]{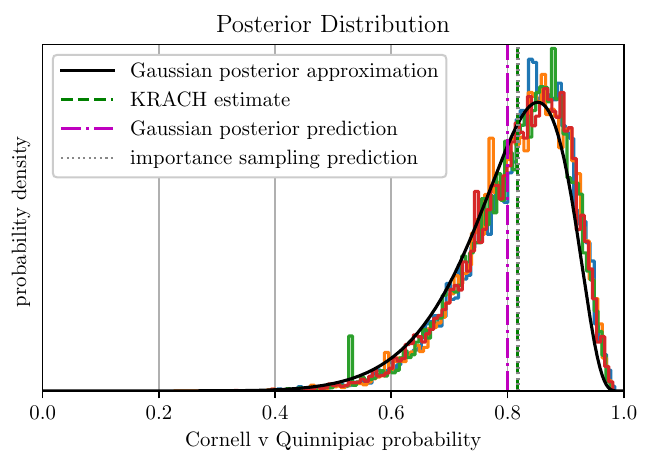}
  \caption{Marginal posterior distribution on
    $\ppair_{\text{Cr}\text{Qn}}$, the probability for Cornell to
    defeat Quinnipiac using the Haldane prior.  Note that the maximum
    likelihood estimate $\ML{\ppair}_{\text{Cr}\text{Qn}}$ associated
    with the KRACH ratings is not the maximum of the marginal
    posterior, because of the transformation of the probability
    density function.  The theoretical curve uses the Gaussian
    approximation, and the histograms are four replications of the
    marginal posterior estimated using importance sampling with the
    Gaussian approximation and 20,000 samples each, as described in
    \sref{s:applications-matrix-demo}}
  \label{f:CrQnprob}
  \label{f:gaussimpCrQnprob}
\end{figure}
As demonstrated in \sref{s:applications-bayes}, the KRACH/MLE
Bradley-Terry model produces reasonably accurate predictions when
applied late in the college hockey season, based on the use of the
model to assign probabilities to NCAA tournament outcomes.  However,
it can lead to some potentially inaccurate probabilities, in
particular in underestimating the probabilities of unlikely events or
sequences of events.  As an illustration, we consider the situation on
2018 March 9, when Cornell and Quinnipiac began a best-of-three
playoff series.  Their respective KRACH ratings were $\elfCrKRACH$ and
$\elfQnKRACH$, so the estimated probability of Cornell winning the game was
$81.7\%$.  However, there was still some uncertainty in the difference
of their Bradley-Terry log-strengths, as illustrated in
\fref{f:CrQnprob}.  Applying the Gaussian approximation of
\sref{s:probabilities-gauss}, we get posterior predictive probability
for Cornell to defeat Quinnipiac of
\begin{equation}
  \int_0^1 \ppair_{\text{Cr}\text{Qn}}\,\pdf(\ppair_{\text{Cr}\text{Qn}}|D,I)
  \,d\ppair_{\text{Cr}\text{Qn}}
  \approx\elfCrQngausspct\%
\end{equation}

\begin{figure}[t!]
  \centering
  \includegraphics[width=\textwidth]{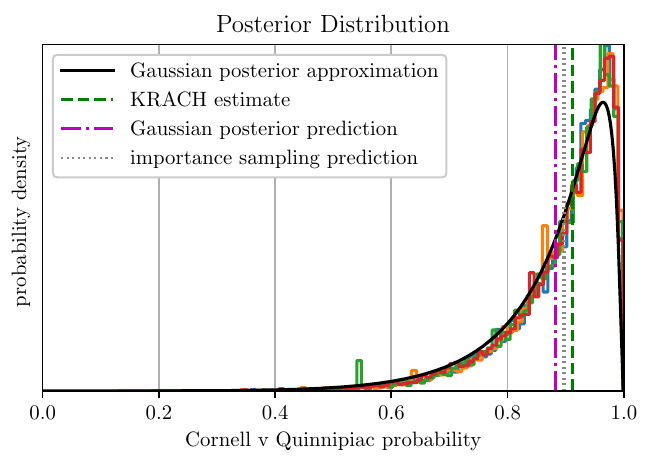}
  \caption{Marginal posterior distribution on the probability
    $\ppair_{\text{Cr}\text{Qn}}^2
    + 2(1-\ppair_{\text{Cr}\text{Qn}})\ppair_{\text{Cr}\text{Qn}}^2$
    for Cornell to defeat Quinnipiac in a best-of-three series, using
    the Haldane prior.  The theoretical curve uses the Gaussian
    approximation, and the histograms are four replications of the
    marginal posterior estimated using importance sampling with the
    Gaussian approximation and 20,000 samples each, as described in
    \sref{s:applications-matrix-demo}}
  \label{f:CrQnseries}
  \label{f:gaussimpCrQnseries}
\end{figure}
An additional effect of the uncertainty is that the game results are
not independent.  If Cornell lost one game with Quinnipiac, it would
mean the difference of their Bradley-Terry log-strengths was more
likely to be below $\ML{\lteam}_{\text{Cr}}-\ML{\lteam}_{\text{Qn}}$
than above
it, and the adjusted posterior probability they would lose another
game would be higher.\footnote{In practice, one calculates the
  probability for the whole sequence of results, but it can be
  conceptually understood according to the Bayesian updating of
  posteriors, where
  $\prob(O_1,O_2|D,I)=\prob(O_2|O_1,D,I)\prob(O_1|D,I)$.} This is
reflected in \fref{f:CrQnseries}, which shows that while the
probabilities from the KRACH ratings would give Cornell a
{\elfCrQnKRACHseriespct}\%
chance to win two out of three games with Quinnipiac, the actual
posterior predictive probability (using the Gaussian approximation) is
\begin{equation}
  \int_0^1
  \left[
    \ppair_{\text{Cr}\text{Qn}}^2
    + 2(1-\ppair_{\text{Cr}\text{Qn}})\ppair_{\text{Cr}\text{Qn}}^2
  \right]
  \,
  \pdf(\ppair_{\text{Cr}\text{Qn}}|D,I)
  \,d\ppair_{\text{Cr}\text{Qn}}
\approx\elfCrQngaussseriespct\%
\end{equation}

\subsubsection{Proposed Modification to the Pairwise Probability
  Matrix}

\label{s:applications-matrix-modification}

Re-calculating the KRACH ratings to account for each simulated game
result would address the correlations between game results, but not
the uncertainties arising from the asymmetries of the marginal
distributions for probabilities like $\ppair_{\text{Cr}\text{Qn}}$.
It would also be rather computationally intensive.  Instead, we
propose to modify the Monte Carlo algorithm to improve the estimate of
probabilities using the Gaussian approximation of
\sref{s:probabilities-gauss} or the importance sampling method of
\sref{s:probabilities-importance}.  The modified Monte Carlo workflow
would be
\begin{enumerate}
\item The multivariate Gaussian approximation is constructed to the
  posterior distribution from the Bradley-Terry log-strengths using
  the Haldane prior with the constraint $\sum_{i=1}^{\nt} \lteam_i=0$;
  the peak is at the maximum-likelihood point $\{\ML{\lteam_i}\}$ and
  the variance-covariance matrix is the pseudo-inverse
  $\{\Sigma_{ij}\}$ of the Hessian matrix
  $H_{ij} = -\nnum_{ij}\ML{\ppair}_{ij}\ML{\ppair}_{ji} +
  \delta_{ij}\sum_{k=1}^{\nt}\ML{\ppair}_{ik}\ML{\ppair}_{ki}$.
\item A set of $N=20,\!000$ Monte Carlo trials are run.  In each trial:
  \begin{enumerate}
  \item A random draw $\{\lteam^{(s)}_{i}|i=1,\ldots,\nt\}$ is made
    from the multivariate normal
    $N_{\nt}(\{\MP{\lteam}_i\},\{\Sigma_{ij}\})$
  \item If importance sampling is to be used, the ratio
    $w_s \propto
    \frac{\pdf(\{\lteam_i^{(s)}\}|D,I)}{g(\{\lteam_i^{(s)}\}|D,I)}$
    of the exact posterior to the sampling distribution, at the point
    $\{\lteam^{(s)}_{i}\}$, is recorded.
  \item The games are simulated using a win probability matrix\\
    $\theta^{(s)}_{ij}=\logistic(\lteam^{(s)}_i-\lteam^{(s)}_j)$ as in
    the current algorithm.
  \item The sequence of games and NCAA selection criteria are created
    from the series of game results as they are now.
  \end{enumerate}
\item If importance sampling is not used, the probability of an
  outcome $O$ is approximated as the fraction of Monte Carlo
  simulations in which it occurs
  \begin{equation}
    \prob(O|D,I)
    \approx \frac{1}{N}\sum_{s=1}^N I^{(s)}(O)
  \end{equation}
  If importance sampling is used, the outcomes are weighted by the
  ratio $w_s$, normalized such that $\sum_{s=1}^N w_s=1$:
  \begin{equation}
    \prob(O|D,I)
    \approx \frac{\sum_{s=1}^N w_s\,I^{(s)}(O)}{\sum_{s=1}^N w_s}
    = \sum_{s=1}^N w_s\,I^{(s)}(O)
  \end{equation}
\end{enumerate}

\subsubsection{Demonstration of Modifications}

\label{s:applications-matrix-demo}

The modifications proposed in \\
\sref{s:applications-matrix-modification} have not yet been
integrated into the generation of the Pairwise Probability Matrix.
However, we can demonstrate their impact by recomputing the
probabilities shown in \sref{s:applications-matrix-CrQn}.  Using the
game results of the 2018-2019 season prior to 2018 March 9, we
construct the Gaussian approximation $g(\{\lteam_i\}|D,I)$ to the
posterior $\pdf(\{\lteam_i\}|D,I)$.  We then draw $N=20,\!000$ samples
$\{\lteam_i^{(s)}\}$ from this distribution, and calculate the weights
$w_s\propto\frac{\pdf(\{\lteam_i^{(s)}\}|D,I)}{g(\{\lteam_i^{(s)}\}|D,I)}$.
For each sample, we have a probability
$\ppair_{\text{Cr}\text{Qn}}^{(s)}=\logistic(\lteam^{(s)}_{\text{Cr}}-\lteam^{(s)}_{\text{Wn}})$
that Cornell will defeat Quinnipiac in a game, and a probability
\begin{equation}
  \pi_{\text{Cr}\text{Qn}}^{(s)}
  = (\ppair_{\text{Cr}\text{Qn}}^{(s)})^2
  + 2 (1-\ppair_{\text{Cr}\text{Qn}}^{(s)})(\ppair_{\text{Cr}\text{Qn}}^{(s)})^2
\end{equation}
that Cornell will win a three-game series.

We simulate a subset of the Pairwise Probability Matrix Monte Carlo as
follows.  For each sample $s$ we make three draws from a Bernoulli
distribution with probability $\ppair_{\text{Cr}\text{Qn}}^{(s)}$.  If
the first draw, $W_{\text{Cr}\text{Qn}}^{(s)}$, is one, we assign that
sample as a win for Cornell; if it is zero, we assign that as a loss
for Cornell.  If two or more of the three draws for a sample are one,
we set $S_{\text{Cr}\text{Qn}}^{(s)}=1$ (series win for Cornell);
otherwise we set $S_{\text{Cr}\text{Qn}}^{(s)}=0$ (series loss for
Cornell).

\begin{table}[t!]
  \label{t:simsCrQn}
  \centering
  \caption{Results of simulations using the Gaussian approximation
    with and without importance sampling to estimate the probability,
    expressed as a percentage, of Cornell winning a game, or a
    best-of-three series, with Quinnipiac, as of 2018 March 9.}
  \begin{tabular}{l|c|c|c}
    \multicolumn{4}{c}{Game (KRACH probability = $\elfCrQnKRACHpct\%$)}
    \\
    \hline
    & integration & MC integration & MC simulation
    \\
    \hline
    Gaussian approx
    & $\elfCrQngausspct$
                  & $\elfCrQnavgpcta$, $\elfCrQnavgpctb$,
                    $\elfCrQnavgpctc$, $\elfCrQnavgpctd$
                                   & $\elfCrQnmcpcta$, $\elfCrQnmcpctb$,
                                     $\elfCrQnmcpctc$, $\elfCrQnmcpctd$
    \\
    \hline
    \multicolumn{2}{c|}{Importance sampling}
    & $\elfCrQnavgimppcta$, $\elfCrQnavgimppctb$,
      $\elfCrQnavgimppctc$, $\elfCrQnavgimppctd$
                  & $\elfCrQnmcimppcta$, $\elfCrQnmcimppctb$,
                    $\elfCrQnmcimppctc$, $\elfCrQnmcimppctd$
  \end{tabular}
  \begin{tabular}{l|c|c|c}
    \multicolumn{4}{c}{Series (KRACH probability = $\elfCrQnKRACHseriespct\%$)}
    \\
    \hline
    & integration & MC integration & MC simulation
    \\
    \hline
    Gaussian approx
    & $\elfCrQngaussseriespct$
                  & $\elfCrQnavgseriespcta$, $\elfCrQnavgseriespctb$,
                    $\elfCrQnavgseriespctc$, $\elfCrQnavgseriespctd$
                                   & $\elfCrQnmcseriespcta$, $\elfCrQnmcseriespctb$,
                                     $\elfCrQnmcseriespctc$, $\elfCrQnmcseriespctd$
    \\
    \hline
    \multicolumn{2}{c|}{Importance sampling}
    & $\elfCrQnavgimpseriespcta$, $\elfCrQnavgimpseriespctb$,
      $\elfCrQnavgimpseriespctc$, $\elfCrQnavgimpseriespctd$
                  & $\elfCrQnmcimpseriespcta$, $\elfCrQnmcimpseriespctb$,
                    $\elfCrQnmcimpseriespctc$, $\elfCrQnmcimpseriespctd$
  \end{tabular}
\end{table}
To estimate the probability of Cornell winning a single game against
Quinnipiac, assuming the Gaussian approximation (which was
analytically computed in \sref{s:applications-matrix-CrQn} to be
$\elfCrQngausspct\%$ by numerical integration of the marginal Gaussian
posterior on $\lteam_{\text{Cr}}-\lteam_{\text{Qn}}$), we can perform
two different calculations:
A Monte Carlo average
  $\frac{1}{N}\sum_{s=1}^N \ppair_{\text{Cr}\text{Qn}}^{(s)}$ of the
  single-game probability,
  or
the fraction
  $\frac{1}{N}\sum_{s=1}^N W_{\text{Cr}\text{Qn}}^{(s)}$ of the of
  simulations in which Cornell wins the first game.  The latter is the
  analogue of what would be computed in the Pairwise Probability
  Matrix.
To adjust the single-game computation using importance sampling, we
again have two options:
A weighted Monte Carlo average
  $\sum_{s=1}^N w_s\,\ppair_{\text{Cr}\text{Qn}}^{(s)}$ of the
  single-game probability,
  or
the sum of the weights
  $\sum_{s=1}^N w_s\,W_{\text{Cr}\text{Qn}}^{(s)}$ of the of
  simulations in which Cornell wins the first game.  The latter is the
  analogue of what would be computed in the Pairwise Probability
  Matrix.
To test these, and estimate Monte Carlo errors, we performed four
replications of the whole process (with 20,000 Monte Carlo samples
each).  The histograms of the simulated probabilities
$\ppair_{\text{Cr}\text{Qn}}^{(s)}$, weighted by $w_s$, are plotted in
\fref{f:gaussimpCrQnprob}.  We see that there are some
heavily-weighted outliers (the largest weights in the four
replications are $\elfmaxweighta$, $\elfmaxweightb$, $\elfmaxweightc$,
and $\elfmaxweightd$, compared to an average weight of $0.00005$).
However, when we estimate the single-game probabilities using
importance sampling, summarized in \tref{t:simsCrQn}, they all come
out consistently, slightly below 82\%, and we can distinguish a small
difference between the probability predicted with and without
importance sampling.  This is also reflected in the histograms in
\fref{f:gaussimpCrQnprob}, where we can see that, despite the
outliers, the overall shape of the estimated posterior appears to be
skewed a bit further right than the one derived from the Gaussian
approximation on $\{\lteam_i\}$.  Note that the effects of including
the posterior uncertainty, and going from the Gaussian approximation
to the posterior estimated by importance sampling, cancel out, and the
estimated probability is quite close to that calculated from the
maximum-likelihood/KRACH approximation.  We will see when we consider
three-game series that this cancellation is a coincidence.

To estimate the probability of Cornell winning a best-of-three series
against Quinnipiac, assuming the Gaussian approximation (which was
computed in\\
\sref{s:applications-matrix-CrQn} using numerical
integration as $\elfCrQngaussseriespct\%$), we can again perform two
calculations:
A Monte Carlo average
  $\frac{1}{N}\sum_{s=1}^N \pi_{\text{Cr}\text{Qn}}^{(s)}$ of the
  single-game probability,
  or
the fraction
  $\frac{1}{N}\sum_{s=1}^N S_{\text{Cr}\text{Qn}}^{(s)}$ of the of
  simulations in which Cornell wins the first game.  The latter is the
  analogue of what would be computed in the Pairwise Probability
  Matrix.
Similarly, when we use importance sampling, we can compute either
a weighted Monte Carlo average
  $\sum_{s=1}^N w_s\,\pi_{\text{Cr}\text{Qn}}^{(s)}$ of the
  best-of-three probability,
  or 
the sum of the weights
  $\sum_{s=1}^N w_s\,S_{\text{Cr}\text{Qn}}^{(s)}$ of the of
  simulations in which Cornell wins the three-game series.  The latter is the
  analogue of what would be computed in the Pairwise Probability
  Matrix.
The histograms of the simulated probabilities
$\pi_{\text{Cr}\text{Qn}}^{(s)}$, weighted by $w_s$, are plotted in
\fref{f:gaussimpCrQnseries}, and the probabilities are summarized in
\tref{t:simsCrQn}.  We see that the Monte Carlo simulations of the
games are somewhat more robust than the Monte Carlo averages, but the
results are consistent, just below 90\%, and noticeably different from
both the Gaussian approximation (about 88\%) and the maximum
likelihood/KRACH probability of 91.1\%.

\section{Discussion}

We have illustrated some applications of the Bradley-Terry model to
college hockey.  The model, in its maximum-likelihood form, is already
used to rank teams as the basis of the KRACH ratings.  Because the
Bradley-Terry strength parameters naturally produce probabilities of
game outcomes, the model can also be used for the prediction of future
outcomes based on past results.  In \sref{s:probabilities} we showed
how to go beyond the maximum-likelihood values of these probabilities
to account for posterior uncertainties in the parameters and estimate
posterior predictive probabilities.  One can avoid the use of full
Markov Chain Monte Carlo methods by approximating the relevant
marginalization integrals using a multivariate Gaussian approximation
to the posterior and/or importance sampling.

In \sref{s:applications} we exhibited two applications of these
posterior predictive probabilities.  The first used these probabilities
to evaluate the models (Bradley-Terry or otherwise) generating them.
Constructing a Bayes factor for the NCAA tournament results using the
probabilities predicted using the pre-tournament results shows, over
time, the superiority of Bradley-Terry models to a na\"{\i}ve model
based only on each team's win/loss ratio.  It would be illuminating to
compare the Bradley-Terry model to more sophisticated alternatives
such as the Ratings Percentage Index (RPI), but that is non-trivial
because the RPI doesn't naturally produce predictive probabilities.

Finally, the Pairwise Probability Matrix is a natural application of
the Bradley-Terry model to assign probabilities to the outcome of the
last few weeks of a college hockey season, in terms of which teams
qualify for the NCAA tournament.  The current application uses Monte
Carlo simulation to estimate these probabilities from the
maximum-likelihood Bradley-Terry parameters (KRACH ratings).  We have
proposed a modification to this program where, at each Monte Carlo
iteration the ratings are also randomly drawn from an approximation to
their posterior distribution.  This should more accurately account for
posterior uncertainties in the parameters and induced correlations of
future game outcomes.

\bibliographystyle{acm}

\begin{acknowledgment}
  JTW wishes to thank Kenneth Butler, Gabriel Phelan, the attendees of
  the UP-STAT conferences, and the members of the Schwerpunkt
  Stochastik at Goethe University, Frankfurt am Main, for useful
  discussions.  Parts of \sref{s:applications-matrix-CrQn} were
  inspired by a discussion on the eLynah forum.  Game results for the
  computations in this paper were collected from \\
  \texttt{https://www.collegehockeynews.com/schedules/composite.php}

  The python code used to perform the simulations in this paper is
  available at \texttt{https://gitlab.com/jtwsma/bradley-terry}
\end{acknowledgment}

\end{document}